\documentclass[11pt,twoside]{article}
\usepackage{asp2010}

\begin{document}


\title{Morpho-kinematic Modeling of Nova Ejecta}
\author{Val\'erio A. R. M. Ribeiro
\affil{Astrophysics, Cosmology and Gravity Centre, Department of Astronomy, University of Cape Town, Private Bag X3, Rondebosch, 7701, South Africa; vribeiro@ast.uct.ac.za}}


\begin{abstract}
Morpho-kinematic modeling allows us to disentangle the morphology and kinematics of an object. The technique has been applied to a number of novae where resolved imaging, or lack of, and spectroscopic line profile fitting, show how we may retrieve important parameters of the system, such as the maximum expansion velocity, inclination angle and the morphology of the ejected shell. Furthermore, this technique may be used as a predictor for searches of eclipses which will then provide us further information on the system parameters, such as the orbital period and the white dwarf mass.
\end{abstract}

\section{Introduction}
Nova ejecta have been resolved in the optical with a myriad of structures \citep*{H72,S83,SOD95,GO00,HO03}. The most widely accepted mechanism for the formation of these structures is that of a common envelope phase where the ejecta engulfs the secondary star within a matter of minutes. The secondary then transfers energy and angular momentum to the ejecta (for a recent review see \citealt{OB08}). However, a new mechanism has been put forward where the mass loss from the secondary, during quiescence, is highly concentrated in the orbital plane producing naturally the bipolar structures of  the ejecta, with possibly an equatorial waist \citep*{MP12,MBP13}.

Morpho-kinematic modeling\footnote{Using {\sc shape} \citep{SKW11}, available from \url{http://bufadora.astrosen.unam.mx/shape/}} involves disentangling the morphology and kinematics of an object providing information on the expansion velocity ($V_{\rm exp}$), inclination angle and the morphology of the ejected shell following a nova outburst. In this proceeding, some examples are shown of the potential of this technique in retrieving these key parameters. In Section 2, the potential of applying this technique to both resolved imaging and ground-based spectroscopic observations is demonstrated and in Section 3 a particular case where only the optical spectroscopic emission line profile were available is presented. Finally, in Section 4 a discussion of the implications of these results are presented.

\section{Modeling with Resolved Imaging and Spectroscopic Observations}
\begin{figure}[t]
\centering
\includegraphics[scale=0.35]{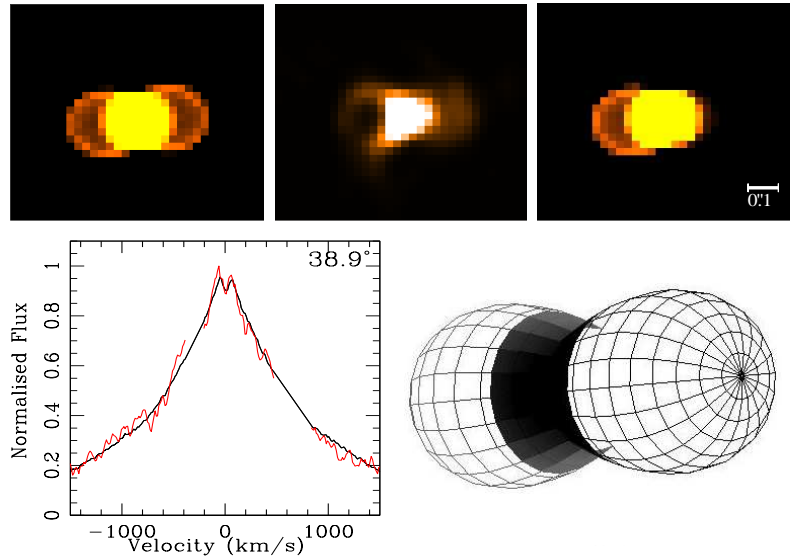}
\caption{Top: synthetic image without the {\it HST} F502N ACS/HRC filter profile applied (left), enlarged ACS/HRC image at $t$ = 155 days after outburst (middle) and synthetic image with the ACS/HRC F502N filter profile applied (right). Bottom: best fit synthetic spectrum (black) overlaid with the observed spectrum (red). To the right is the model structure for RS Oph (outer dumbbell and inner hour glass). Images reproduced from \citet{RBD09}.
}
\label{fig1}
\end{figure}

{\it Hubble Space Telescope} ({\it HST}) ACS/HRC narrow band (F502N filter) imaging and ground-based spectroscopic observations of the recurrent nova RS Ophiuch at 155 days after outburst allowed \citet{RBD09} to model the ejecta as a bipolar composed of an outer dumbbell and inner hour glass structures (Figure~\ref{fig1}). The inner hour glass was required as an over-density in order to replicate the observed [O~{\sc iii}] 5007\AA\ line profile. The apparent asymmetry of the ejecta in the {\it HST} image was shown to be an observational effect due to the finite width and offset of the central wavelength from the line centre of the F502N filter. From detailed kinematic modeling, the inclination of the system was determined as 39$^{+1}_{-9}$ degrees and maximum expansion velocity 5100$^{+1500}_{-100}$~km~s$^{-1}$ (the range in velocity arises from the 1$\sigma$ errors on the inclination, Figure~\ref{fig1}). This asymmetry was proposed to be due to interaction of the ejecta with a pre-existing red-giant wind \citep{BHO07,RBD09}.

The model was then evolved to 449 days after outburst, to match the second {\it HST} observations. However, in this case, due to the lack of simultaneous ground-based spectroscopy the model was harder to constrain. \citet{RBD09} suggested that at this time the outer dumbbell structure expanded linearly while the inner hour glass structure showed some evidence for deceleration. It was not until \citet*{RBW13a} performed archival searches of ground based facilities that they found spectroscopic data on day 415 which suggest the interpretation by \citet{RBD09} may not be far from the truth \citep*[see also][for an update]{RBW13}.

\section{Line Profile Fitting}
\begin{figure}
\centering
\includegraphics[scale=0.15]{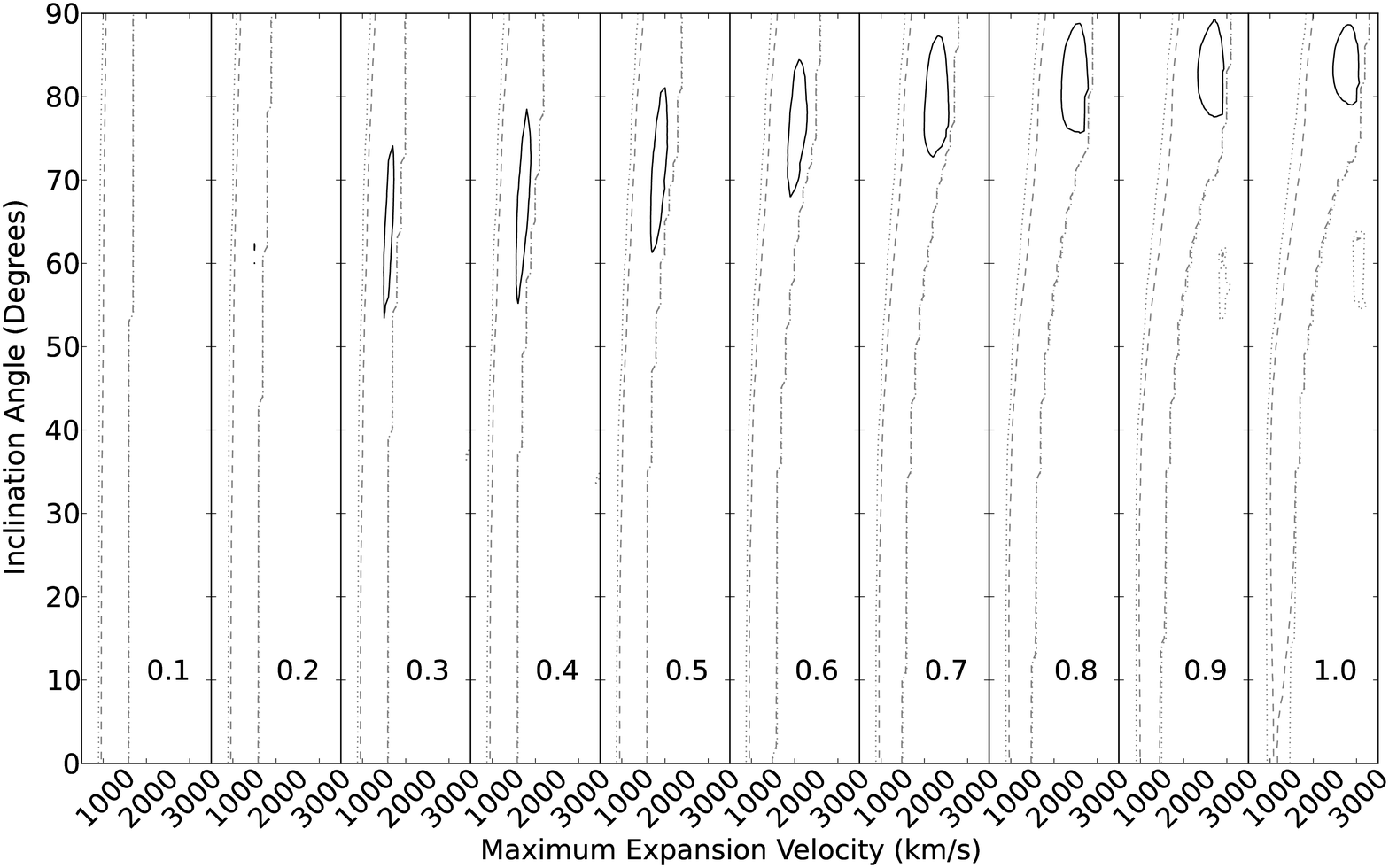} 
\includegraphics[scale=0.25]{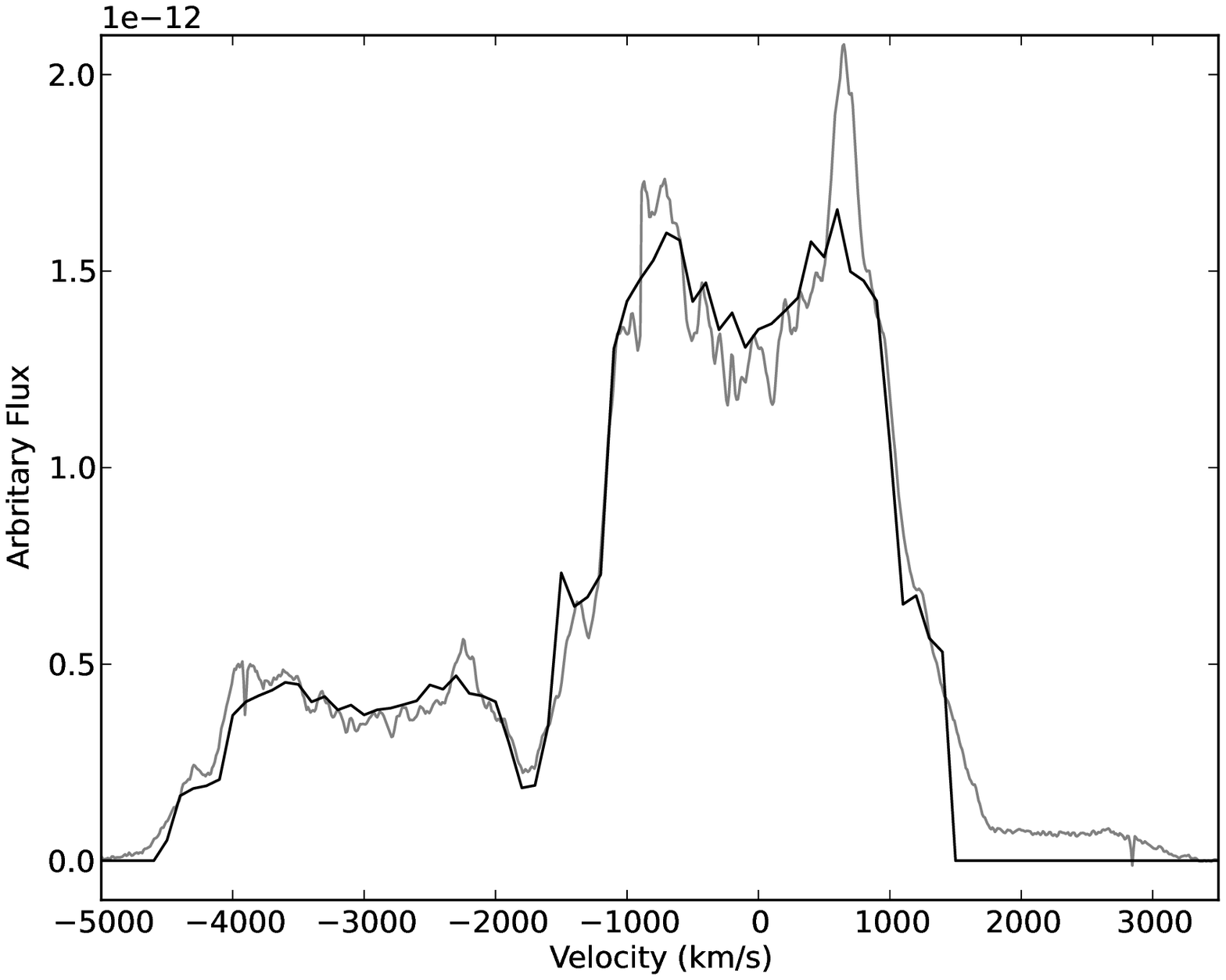}
\caption{Nova Mon 2012 best fit models. $Top$ -- probability density function results for different degrees of bipolarity (black contour represents PDF = 1$\sigma$, gray dashed = 0.05 and dotted = 0.01). $Bottom$ -- best fit model (black) to the observed day 130 after outburst (gray) for a bipolar structure with a pinch of 0.8, derived as one minus the ratio of the semi-major to the semi-minor axis of the ejected shell. Images from \citet{RMV13}.}
\label{fig2}
\end{figure}

Unfortunately, nature does not always provide such good quality data as described above. In subsequent work, only novae with spectroscopic observations were available. In most cases for example, V2672 Oph \citep{MRB11}, V2491 Cyg \citep{RDB11} and KT Eri \citep{R11,RBD13} only the H$\alpha$ emission line was reproduced, due to lack the of forbidden lines. Nova Mon 2012 showed forbidden lines and fitting these lines suggested a bipolar morphology with an inclination angle of 82$\pm$6 degrees and  $V_{\rm exp}$ = 2400$^{+300}_{-200}$ km s$^{-1}$ \citep*[Figure~\ref{fig2},][]{RMV13}. A 7.1~hr periodicity was found in the X-rays \citep{POW13} and shown to be near sinusoidal in shape with a weak secondary minimum \citep{MCD13}.

\section{Discussion}
Ideally to constrain a model, the interplay between resolved imaging and ground based spectroscopic observations play a great role. However, this is not always the case. Applying different models based on observations of novae can aid in constraining the best morphology and hence, the inclination angle and maximum expansion velocity.

The success of modeling RS Oph, suggested the existence of an extra component that was required to determine the observed low velocity of the emission line profile while the imaging showed much higher velocity material. This extra component has been suggested to be material that survives the outburst \citep{EWH07,RBW13a}. Further study is required here.

Of particular interest to Nova Mon 2012, besides its $\gamma$--ray origin, is the e-VLBI observations of two components which may be associated with the ejecta \citep{OYP12}. Furthermore, resolved VLBA imaging showed what appeared to be a bipolar morphology of Nova Mon 2012 (M. Rupen, private communication). The derived inclination angle suggest eclipses should be observed. In fact, at least in Nova Mon 2012, a orbital period of 7.1~hr has been observed which was suggested to be due to partial eclipse from extended emission by an accretion disk rim \citep{POW13}. \citet{MCD13} showed the light curve to have a near sinusoidal shape with a weak secondary minima at phase 0.5. This was interpreted as arising from the super-imposed ellipsoidal distortion of the K3~V Roche lobe filling secondary and irradiation of its side facing the WD. All these results agree well with the high inclination and bipolar morphology derived by \citet{RMV13}.

It is not always that we have great success with deriving the best-fit model. In some cases the model is not as simple as just bipolar. For example, in V2491 Cyg the best-fit model structure was that of polar blobs and an equatorial ring. The model was allowed to evolve from the early outburst phase to later stages. This evolution reproduced well the observed spectra, in particular when [{N {\sc ii}}], in either side of the H$\alpha$ line, also modeled \citep{RDB11}.

\acknowledgements The author is grateful to long term collaborators, Mike Bode, Matt Darnley, Dan Harman and Ulisse Munari for insightful discussions over the years and Wolfgang Steffen and Nico Koning for valuable discussion and continued development of {\sc shape}. The author acknowledges the South African SKA Project for funding the postdoctoral fellowship at the University of Cape Town.

\end{document}